%% file: babar-pub-15424.tex
\documentclass[ reprint,superscriptaddress,showpacs, amsmath,amssymb, aps, prl,floatfix,]{revtex4-1}

\usepackage{color}
\usepackage{graphicx}
\usepackage{dcolumn}% Align table columns on decimal point
\usepackage{bm}% bold math
\usepackage{subfigure}
\usepackage{multirow}
\newcommand{\BABARPubYear}       {13}
\newcommand{\BABARPubNumber}     {004}
\newcommand{\SLACPubNumber} {15424}

\input{babarsym}

\long\def\inst#1{\par\nobreak\kern 4pt\nobreak
    {\it #1}\par\vskip 10pt plus 3pt minus 3pt}

\begin{document}

\preprint{\babar-PUB-\BABARPubYear/\BABARPubNumber} 
\preprint{SLAC-PUB-\SLACPubNumber}

\begin{flushleft}
\babar-PUB-\BABARPubYear/\BABARPubNumber\\
SLAC-PUB-\SLACPubNumber\\
arXiv:1304.5657\\
\vspace{0.4in}
\end{flushleft}

\title{\boldmath{Measurement of the $D^{*}(2010)^{+}$ meson width and the $D^{*}(2010)^{+} - D^0$ mass difference}}\vspace{0.3in}

\input{authors_feb2013_bad2528}

% ----------- ABSTRACT -----------
\begin{abstract}
We measure the mass difference, $\Delta m_0$, between the $D^{*}(2010)^+$ and the $D^0$ and the natural line width, $\Gamma$, of the transition $D^{*}(2010)^+\to D^0 \pi^+$. The data were recorded with the \babar\ detector at center-of-mass energies at and near the $\Upsilon(4S)$ resonance, and correspond to an integrated luminosity of approximately $477 \invfb$. The $D^0$ is reconstructed in the decay modes $D^0 \to K^-\pi^+$ and $D^0 \to K^-\pi^+\pi^-\pi^+$. For the decay mode $D^0\to K^-\pi^+$ we obtain $\Gamma =  \left(83.4 \pm 1.7 \pm 1.5\right) \kev$ and $\Delta m_0 =  \left(145\,425.6 \pm 0.6 \pm 1.8\right) \kev$, where the quoted errors are statistical and systematic, respectively. For the $D^0\to K^-\pi^+\pi^-\pi^+$ mode we obtain $\Gamma = \left(83.2 \pm 1.5 \pm 2.6\right) \kev$ and $\Delta m_0 =  \left(145\,426.6 \pm 0.5 \pm 2.0\right) \kev$.  The combined measurements yield $\Gamma = \left(83.3 \pm 1.2 \pm 1.4\right) \kev$ and $\Delta m_0 = \left(145\,425.9 \pm 0.4 \pm 1.7\right) \kev$; the width is a factor of approximately 12 times more precise than the previous value, while the mass difference is a factor of approximately 6 times more precise.
\end{abstract}
% ----------- END ABSTRACT -----------

\maketitle

\setcounter{footnote}{0}

The line width of the $D^{*}(2010)^{+}$ ($D^{*+}$) provides a window into a nonperturbative regime of strong interaction physics where the charm quark is the heavier meson constituent~\cite{Becirevic201394, actapolb.30.3849, Guetta2001134}. The line width provides an experimental test of $D$ meson spectroscopic models, and is related to the strong coupling of the $D^*D\pi$ system, $g_{D^* D \pi}$. In the heavy-quark limit, which is not necessarily a good approximation for the charm quark~\cite{PhysRevC.83.025205}, this coupling can be related to the universal coupling of heavy mesons to a pion, $\hat{g}$. Since the decay $B^* \rightarrow B\pi$ is kinematically forbidden, it is not possible to measure the coupling $g_{B^*B\pi}$ directly. However, the $D$ and $B$ systems can be related through $\hat{g}$, allowing the calculation of $g_{B^*B\pi}$, which is needed for a model-independent extraction of $\left|V_{ub}\right|$~\cite{PhysRevD.49.2331,PhysRevLett.95.071802} and which forms one of the larger theoretical uncertainties for the determination of $\left|V_{ub}\right|$~\cite{2009PhRvD79e4507B}.

We study the $D^{*+}\to D^0 \pi^+$ transition, using the $D^0\to K^-\pi^+$ and $D^0\to K^-\pi^+\pi^-\pi^+$ decay modes, to extract values of the $D^{*+}$ width $\Gamma$ and the difference between the $D^{*+}$ and $D^0$ masses $\Delta m_0$. Values are reported in natural units and the use of charge conjugate reactions is implied throughout this paper. The only prior measurement of the width is $\Gamma = \left(96 \pm 4 \pm 22\right) \kev$ by the CLEO collaboration, where the uncertainties are statistical and systematic, respectively~\cite{PhysRevD.65.032003}.  In the present analysis, we use a data sample that is approximately 50 times larger. This allows us to apply restrictive selection criteria to reduce background and to investigate sources of systematic uncertainty with high precision.

To extract $\Gamma$, we fit the distribution of the mass difference between the reconstructed $D^{*+}$ and the $D^0$ masses, $\Delta m$. The signal component is described with a P-wave relativistic Breit-Wigner (RBW) function convolved with a resolution function based on a Geant4 Monte Carlo (MC) simulation of the detector response~\cite{geant4}.  

The full width at half maximum (FWHM) of the RBW line shape ($\approx 100 \kev$) is much less than the FWHM of the almost Gaussian resolution function which describes more than 99\% of the signal ($\approx 300 \kev$).  Therefore, near the peak, the observed FWHM is dominated by the resolution function shape. However, the shapes of the resolution function and the RBW differ far away from the pole position. Starting $(1.5 - 2.0) \mev$ from the pole position, and continuing to $(5 - 10) \mev$ away (depending on the $ D^0 $ decay channel), the RBW tails are much larger. The observed event rates in this region are strongly dominated by the intrinsic line width of the signal, not the signal resolution function or the background rate. We use the very different resolution and RBW shapes, combined with the good signal-to-background rate far from the peak, to measure  $ \Gamma $ precisely~\cite{prdversion}.

This analysis is based on a data set corresponding to an integrated luminosity of approximately 477 fb$^{-1}$ recorded at, and 40 \mev below, the $\Upsilon(4S)$ resonance~\cite{Lees2013203}. The data were collected with the \babar\ detector at the PEP-II asymmetric energy \epem\ collider, located at the SLAC National Accelerator Laboratory. The \babar\ detector is described in detail elsewhere~\cite{ref:babar,ref:nim_update}; we summarize the relevant features below. The momenta of charged particles are measured with a combination of a cylindrical drift chamber (DCH) and a 5-layer silicon vertex tracker (SVT), both operating within the $1.5$ T magnetic field of a superconducting solenoid. Information from a ring-imaging Cherenkov detector is combined with specific ionization $(dE/dx)$ measurements from the SVT and DCH to identify charged kaon and pion candidates. Electrons are identified, and photons measured, with a CsI(Tl) electromagnetic calorimeter. The return yoke of the superconducting coil is instrumented with tracking chambers for the identification of muons.

We remove a large amount of combinatorial and $B$ meson decay background by requiring $D^{*+}$ mesons produced in $e^+e^- \to c \bar{c}$ reactions to exhibit an \epem center-of-mass-frame momentum greater than $3.6 \gev$. The entire decay chain is fit using a kinematic fitter with geometric constraints at the production and decay vertex of the $D^0$ and the additional constraint that the $D^{*+}$ laboratory momentum points back to the luminous region of the event. The pion from $D^{*+}$ decay is referred to as the ``slow pion'' (denoted $\pi_s^+$) because of the limited phase space available in the $D^{*+}$ decay. The selection criteria are chosen to provide a large signal-to-background ratio (S/B), in order to increase the sensitivity to the long signal (RBW) tails in the $\Delta m$ distribution; they are not optimized for statistical significance. The criteria are briefly mentioned here and presented in detail in the archival reference for this analysis~\cite{prdversion}. The resolution in $\Delta m$ is dominated by the resolution of the $\pi_s^+$ momentum, especially the uncertainty of its direction due to Coulomb multiple scattering. We implement criteria to select well-measured pions. We define our acceptance angle to exclude the very-forward region of the detector, where track momenta are not accurately reconstructed, as determined using an independent sample of reconstructed $K_{S}^{0}\to \pi^- \pi^+$ decays. The $K_S^0$ reconstructed mass is observed to vary as a function of the polar angle $\theta$ of the $K_S^{0}$ momentum measured in the laboratory frame with respect to the electron beam axis. To remove contributions from the very-forward region of the detector we reject events with any $D^{*+}$ daughter track for which $\cos \theta > 0.89$; this criterion reduces the final samples by approximately 10\%. 

\begin{figure}[!h]
\centering
\subfigure{\includegraphics[scale=0.39]{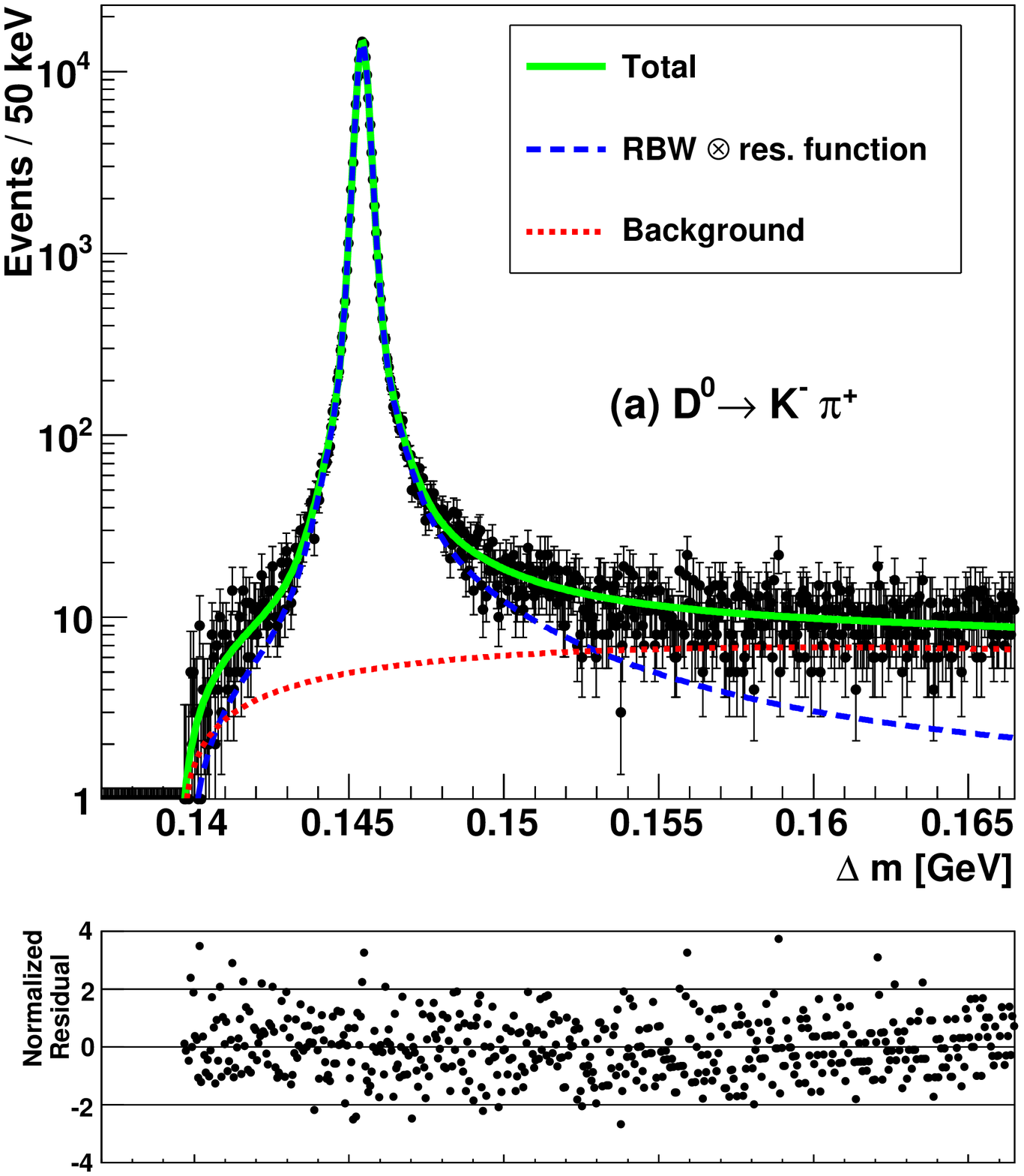}}
\subfigure{\includegraphics[scale=0.39]{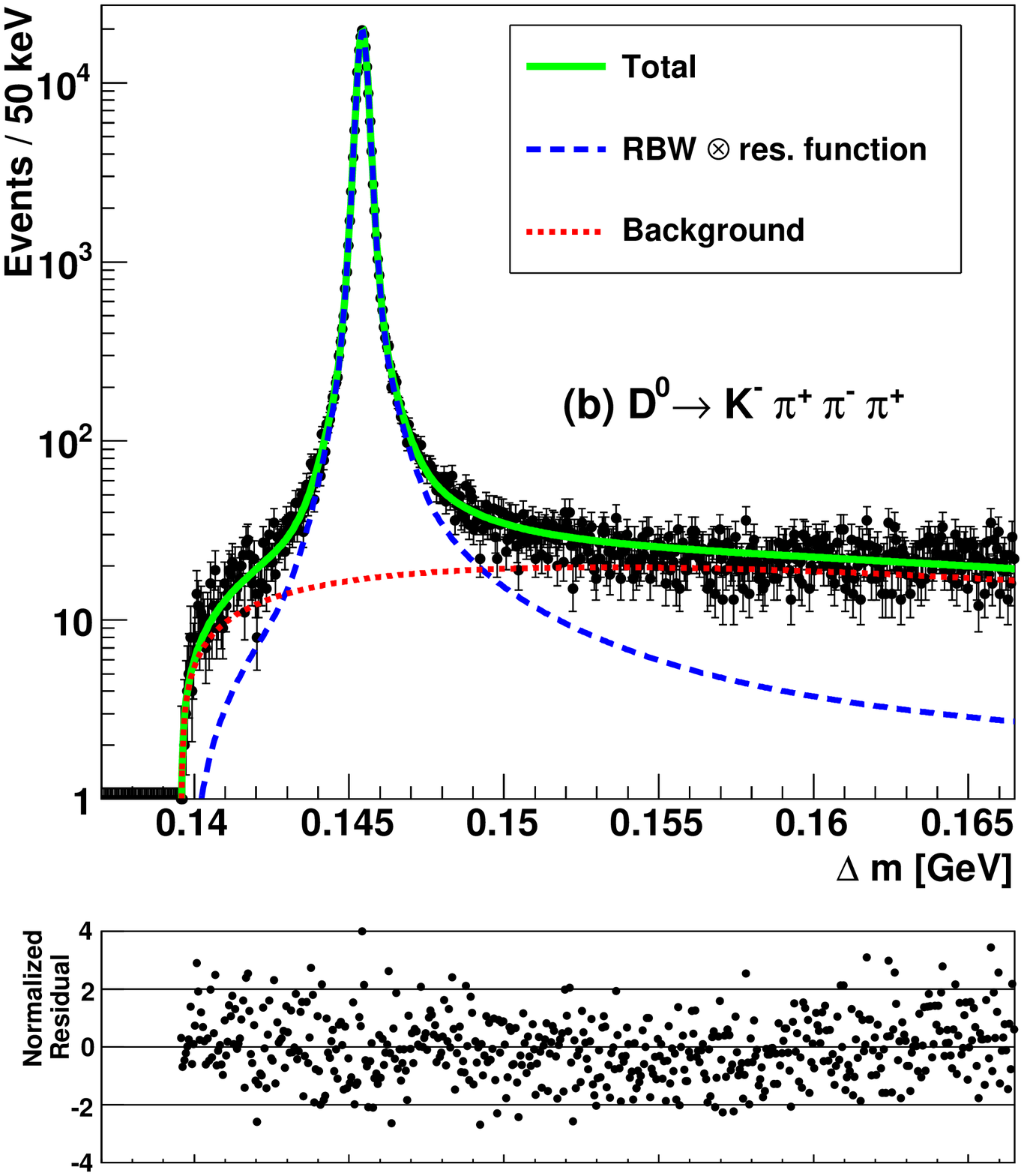}} \\
\caption{(color online) Fits to data for the $D^0\to K^-\pi^+$ and $D^0\to K^-\pi^+\pi^-\pi^+$ decay modes. The total probability density function (PDF) is shown as the solid curve, the convolved RBW-Gaussian signal as the dashed curve, and the background as the dotted curve. The total PDF and signal component are indistinguishable in the peak region. Normalized residuals are defined as $\left(N_{\text{observed}} - N_{\text{predicted}}\right)/\sqrt{N_{\text{predicted}}}$.}
\label{fig:rdfits}
\end{figure}

Our fitting procedure involves two steps. In the first step we model the finite detector resolution associated with track reconstruction by fitting the $\Delta m$ distribution for correctly reconstructed MC events using a sum of three Gaussians and a function to describe the non-Gaussian component~\cite{prdversion}. These simulated $D^{*+}$ decays are generated with $\Gamma = 0.1 \kev$, so that the observed spread of the MC distribution can be attributed to event reconstruction. The non-Gaussian function describes $\pi_s^+$ decays in flight to a $\mu$, for which coordinates from both the $\pi$ and $\mu$ segments are used in track reconstruction. 

The second step uses the resolution shape parameters from the first step and convolves the Gaussian components with a RBW function to fit the measured $\Delta m$ distribution in data. The RBW function is defined by
\begin{equation}
  \frac{d \Gamma(m)}{d m} =  \frac{m \Gamma_{D^*D \pi}\left(m\right) \, m_0 \Gamma}  {\left(m_0^2  - m^2\right)^2  + \left(m_0 \Gamma _{\text{Total}}(m) \right)^2},
  \label{eq:rbw}
\end{equation}
\noindent where $\Gamma_{D^*D \pi}$ is the partial width to $D^0\pi_s^+$, $m$ is the $D^0 \pi_s^+$ invariant mass, $m_0$ is the invariant mass at the pole, $\Gamma_{\text{Total}}(m)$ is the total $D^{*+}$ decay width, and $\Gamma$ is the natural line width we wish to measure. The partial width is defined by 
\begin{equation}
  \Gamma_{D^*D \pi}(m) = \Gamma
  \left(\frac{\mathcal{F}^{\ell}_{D\pi}(p_0)}{\mathcal{F}^{\ell}_{D\pi}(p)}\right)^2\left(\frac{p}{p_0}\right)^{2\ell+1}\left(\frac{m_0}{m}\right).
  \label{eq:partialwidth}
\end{equation}
\noindent Here $\ell = 1$, $\mathcal{F}^{\ell = 1}_{D\pi}\left(p\right) = \sqrt{1+r^2 p^2}$ is a Blatt-Weisskopf form factor for a vector particle with radius parameter $r$ and daughter momentum $p$, and the subscript zero denotes a quantity measured at the mass pole $m_0$ \cite{blatt, PhysRevD.5.624}. We use the value $r = 1.6 \gev^{-1}$ from Ref.~\cite{Schwartz:2002hh}. For the purpose of fitting the $\Delta m$ distribution, we obtain $d \Gamma(\Delta m)/d \Delta m$ from Eqs. (\ref{eq:rbw}) and (\ref{eq:partialwidth}) through the substitution $m = m(D^0) + \Delta m$, where $m(D^0)$ is the nominal $D^0$ mass~\cite{ref:pdg2012}.

As in the CLEO analysis~\cite{PhysRevD.65.032003}, we approximate the total $D^{*+}$ decay width $\Gamma_{\text{Total}}(m) \approx \Gamma_{D^*D \pi}(m)$, ignoring the electromagnetic contribution from $D^{*+}\to D^+ \gamma$. This approximation has a negligible effect on the extracted values, as it appears only in the denominator of the RBW function.

To allow for differences between MC simulation and data, the root-mean-square deviation of each Gaussian component of the resolution function is allowed to scale in the fit process by the common factor $(1 + \epsilon)$. Events that contribute to the non-Gaussian component have a well-understood origin ($\pi_s$ decay in flight), which is accurately reproduced by MC simulation. 
In the fit to data, the non-Gaussian function has a fixed shape and relative fraction, and is not convolved with the RBW.
The relative contribution of the non-Gaussian function is small ($\lesssim0.5$\% of the signal), and the results from fits to validation signal-MC samples are unbiased without convolving this term. The background is described by a phase-space model of continuum background near the kinematic threshold~\cite{prdversion}. We fit the $\Delta m$ distribution from the kinematic threshold to $\Delta m = 0.1665 \gev$ using a binned maximum likelihood fit and an interval width of $50 \kev$.

In the initial fits to data, we observed a strong dependence of $\Delta m_0$ on the slow pion momentum. This dependence, which originates in the modeling of the magnetic field map and the material in the beam pipe and SVT, is not replicated in the simulation. Previous \babar\ analyses have observed similar effects, for example the measurement of the $\Lambda_c^+$ mass~\cite{PhysRevD.72.052006}. In that analysis the material model of the SVT was altered in an attempt to correct for the energy loss and the under-represented small-angle multiple scattering (due to nuclear Coulomb scattering). However, the momentum dependence of the reconstructed $\Lambda_c^+$ mass could be removed only by adding an unphysical amount of material to the SVT. In this analysis we use a different approach to correct the observed momentum dependence and adjust track momenta after reconstruction.

We use a sample of $K_{S}^{0}\to \pi^+\pi^-$ events from $D^{*+} \to D^0 \pi_s^+$ decays, where $D^0 \to K_S^0 \pi^+ \pi^-$, and require that the $K_{S}^{0}$ daughter pions satisfy the same tracking criteria as the $\pi_s^+$ candidates for the $D^0\to K^-\pi^+$ and $D^0\to K^-\pi^+\pi^-\pi^+$ signal modes. The $K_{S}^0$ decay vertex is required to lie inside the beam pipe and to be well separated from the $D^{0}$ vertex.  
These selection criteria yield an extremely clean $K_{S}^0$ sample (over 99.5\% pure), which is use to determine three fractional corrections to the overall magnetic field and to the energy losses in the beam pipe and, separately, in the SVT. We determine the best set of correction parameters by minimizing the difference between the $\pi^+\pi^-$ invariant mass and the current world average for the $K^{0}$ mass ($497.614 \pm 0.024 \mev$)~\cite{ref:pdg2012} in 20 intervals of laboratory momentum in the range $0.0$ to $2.0 \gev$. The best-fit parameters increase the magnitude of the magnetic field by 4.5 Gauss and increase the energy loss in the beam pipe and SVT by 1.8\% and 5.9\%, respectively~\cite{prdversion}.  The momentum-dependence of $\Delta m_0$ in the preliminary results was mostly due to the slow pion. However, the correction is applied to all $D^{*+}$ daughter tracks. All fits to data described in this analysis are performed using masses and $\Delta m$ values calculated using corrected momenta. Simulated events do not require correction because the same field and material models used to propagate tracks are used for their reconstruction.

Figure~\ref{fig:rdfits} presents the results of the fits to data for both the $D^0\to K^-\pi^+$ and $D^0\to K^-\pi^+\pi^-\pi^+$ decay modes. The normalized residuals show good agreement between the data and our fits.  Table~\ref{table:kpik3pi_rd_summary} summarizes the results of the fits to data for both $D^0$ decay modes.  The table also shows S/B at the peak and in the high $\Delta m$ tail of each distribution.

\begin{table}[!h]
\centering
\caption{Summary of the results from the fits to data for the $D^0\to K^-\pi^+$ and $D^0\to K^-\pi^+\pi^-\pi^+$ channels (statistical uncertainties only).  S/B is the ratio of the convolved signal PDF to the background PDF at the given $\Delta m$ and $\nu$ is the number of degrees of freedom.}

\begin{tabular}{c@{\hspace{4mm}}c@{\hspace{4mm}}c}
\hline \hline \\ [-1.7ex]
Parameter & $D^0\to K\pi$ & $D^0\to K\pi\pi\pi$ \\ \hline \\[-1.7ex]
Number of signal events & $138\,536 \pm 383$ &  $174\,297\pm 434$ \\ 
$\Gamma \,(\kev)$ & $83.3 \pm 1.7$ & $83.2 \pm 1.5$ \\ 
scale factor, $(1+\epsilon)$ & $1.06 \pm 0.01$ & $1.08 \pm 0.01$ \\ 
$\Delta m_0\,(\kev)$ & $145\,425.6 \pm 0.6$ & $145\,426.6 \pm 0.5$ \\  \\[-1.7ex]
$S/B$ at peak & \multirow{2}{*}{$2700$} & \multirow{2}{*}{$1130$} \\ 
($\Delta m = 0.14542 \,(\gev)$) & & \\ \\ [-1.8ex]
$S/B$ at tail & \multirow{2}{*}{$0.8$} & \multirow{2}{*}{$0.3$} \\
($\Delta m = 0.1554\, (\gev)$) & & \\ \\ [-1.8ex]
$\chi^2/\nu$ & $574/535$ & $556/535$ \\[-1.7ex] \\ \hline \hline
\end{tabular}
\label{table:kpik3pi_rd_summary}
\end{table}

\begin{table*}[ht]
\centering
\caption{Summary of systematic uncertainties with correlation $\rho$ between the $D^0\to K^-\pi^+$ and $D^0\to K^-\pi^+\pi^-\pi^+$ modes. The $K^-\pi^+$ and $K^-\pi^+\pi^-\pi^+$ invariant masses are denoted by $m\left(D^0_{\text{reco}}\right)$.}
\begin{tabular}{c@{\hspace{8mm}}ccc@{\hspace{8mm}}ccc}
\hline \hline  \\[-1.7ex] 
\multirow{2}{*}{Source} & \multicolumn{2}{c}{$\sigma_{sys} \left(\Gamma\right) \,[\kev]$} & \multirow{2}{*}{$\rho$} & \multicolumn{2}{c}{$\sigma_{sys} \left(\Delta m_0\right) [\kev]$}  & \multirow{2}{*}{$\rho$}\\  \\[-1.7ex] 
& $K\pi$ & $K\pi\pi\pi$ & & $K\pi$ & $K\pi\pi\pi$ & \\ \hline \\[-1.7ex]
Disjoint $D^{*+}$ momentum variation & 0.88 & 0.98 & \phantom{-}0.47 & 0.16 & 0.11 & 0.28 \\ 
Disjoint $m\left(D^0_{\text{reco}}\right)$ variation & 0.00 & 1.53 & \phantom{-}0.56 &  0.00 &  0.00 & 0.22 \\ 
Disjoint azimuthal variation & 0.62 & 0.92 & -0.04 & 1.50 & 1.68 & 0.84 \\ 
Magnetic field and material model & 0.29 & 0.18 & \phantom{-}0.98 & 0.75 & 0.81 & 0.99 \\ 
Blatt-Weisskopf radius & 0.04 & 0.04 & \phantom{-}0.99 & 0.00 & 0.00 & 1.00 \\ 
Variation of resolution shape parameters & 0.41 & 0.37 & \phantom{-}0.00 & 0.17 & 0.16 & 0.00  \\ 
$\Delta m$ fit range & 0.83 & 0.38 & -0.42 &  0.08 & 0.04 & 0.35  \\ 
Background shape near threshold & 0.10  & 0.33  & \phantom{-}1.00 & 0.00 & 0.00 & 0.00  \\ 
Interval width for fit & 0.00 & 0.05 & \phantom{-}0.99 & 0.00 & 0.00 & 0.00 \\ 
Bias from validation & 0.00 & 1.50 & \phantom{-}0.00 & 0.00 & 0.00 & 0.00 \\
Radiative effects & 0.25 & 0.11 & \phantom{-}0.00 & 0.00 & 0.00 & 0.00 \\ \hline \\[-1.7ex]
Total & 1.5 & 2.6 & & 1.7 & 1.9 & \\[-1.7ex]\\ \hline \hline
\end{tabular}
\label{table:syswithcorr}
\end{table*}

We estimate systematic uncertainties related to a variety of sources. The data are divided into disjoint subsets corresponding to intervals of $D^{*+}$ laboratory momentum, $D^{*+}$ laboratory azimuthal angle $\phi$, and reconstructed $D^0$ mass, in order to search for variations larger than those expected from statistical fluctuations. These are evaluated using a method similar to the PDG scale factor~\cite{prdversion, ref:pdg2012}. The corrections to the overall momentum scale and $dE/dx$ loss in detector material are varied to account for the uncertainty on the $K_{S}^{0}$ mass. To estimate the uncertainty in the Blatt-Weisskopf radius we model the $D^{*+}$ as a point-like particle. We vary the parameters of the resolution function according to the covariance matrix reported by the fit to estimate systematic uncertainty of the resolution shape. We vary the end point used in the fit, which affects whether events are assigned to the signal or background component. This variation allows us to evaluate a systematic uncertainty associated with the background parametrization; within this systematic uncertainty, the residual plots shown in Fig.~\ref{fig:rdfits} are consistent with being entirely flat. Additionally, we vary the description of the background distribution near threshold. We fit MC validation samples to estimate systematic uncertainties associated with possible biases. Finally, we use additional MC validation studies to estimate possible systematic uncertainties due to radiative effects. All these uncertainties are estimated independently for the $D^0\to K^-\pi^+$ and $D^0\to K^-\pi^+\pi^-\pi^+$ modes, as discussed in detail in Ref.~\cite{prdversion} and summarized in Table~\ref{table:syswithcorr}. 

The largest systematic uncertainty arises from an observed sinusoidal dependence for $\Delta m_0$ on $\phi$. Variations with the same signs and phases are seen for the reconstructed $D^0$ mass in both $D^0 \to K^-\pi^+$, $D^0\to K^-\pi^+\pi^-\pi^+$, and for the $K_{S}^{0}$ mass. An extended investigation revealed that at least part of this dependence originates from small errors in the magnetic field from the map used in track reconstruction~\cite{prdversion}. The important aspect for this analysis is that the average value is unbiased by the variation in $\phi$, which we verified using the reconstructed $K_S^0$ mass value. The width does not display a $\phi$ dependence, but each mode is assigned a small uncertainty because some deviations from uniformity are observed. The lack of a systematic variation of $\Gamma$ with respect to $\phi$ is notable because $\Delta m_0$ shows a clear dependence such that the results from the $D^0\to K^-\pi^+$ and $D^0\to K^-\pi^+\pi^-\pi^+$ samples are highly correlated and shift together.  We fit the $\Delta m_0$ values with a sinusoidal function and take half of the amplitude as the estimate of the uncertainty.

The results for the two independent $ D^0 $ decay modes agree within their uncertainties.  The dominant systematic uncertainty on the RBW pole position comes from the variation in $\phi$ ($1.5-1.9 \kev$).  For the decay mode $D^0\to K^-\pi^+$ we find $\Gamma = \left(83.4 \pm 1.7 \pm 1.5\right) \kev$ and $\Delta m_0 =  \left(145\,425.6 \pm 0.6 \pm 1.8\right) \kev$, while for the decay mode $D^0\to K^-\pi^+\pi^-\pi^+$ we find $\Gamma = \left(83.2 \pm 1.5 \pm 2.6\right) \kev$ and $\Delta m_0 = \left(145\,426.6 \pm 0.5 \pm 2.0\right) \kev$. Accounting for correlations, we obtain the combined measurement values $\Gamma = \left(83.3 \pm 1.2 \pm 1.4\right) \kev$ and \mbox{$\Delta m_0 = \left(145\,425.9 \pm 0.4 \pm 1.7\right) \kev$}.

Using the relationship between the width and the coupling constant~\cite{prdversion}, we can determine the experimental value of $g_{D^{*+}D^0\pi^+}$. Using $\Gamma$ and the masses from Ref.~\cite{ref:pdg2012} we determine the experimental coupling $g_{D^{*+}D^0\pi^+}^{\text{exp}} = 16.92 \pm 0.13 \pm 0.14$, where we have ignored the electromagnetic contribution from $D^{*+} \to D^+ \gamma$. The universal coupling is directly related to $g_{D^*D\pi}$ by $\hat{g} = g_{D^{*+}D^0\pi^+} f_\pi / \left(2 \sqrt{m_{D^0}m_{D^{*+}}}\right)$. This parametrization is different from that used by CLEO~\cite{PhysRevD.65.032003}; it is chosen to match a common convention in the context of chiral perturbation theory, as in Refs.~\cite{PhysRevC.83.025205, PhysRevD.66.074504}. With this relation and $f_\pi = 130.41 \mev$, we find $\hat{g}^{\text{exp}} = 0.570 \pm 0.004 \pm 0.005$. 

Di Pierro and Eichten~\cite{PhysRevD.64.114004} present results in terms of $R$, the ratio of the width of a given state to the universal coupling constant. At the time of their publication, $\hat{g} = 0.82 \pm 0.09$ was consistent with the values from all of the modes in Ref.~\cite{PhysRevD.64.114004}. In 2010, \babar\ published much more precise mass and width results for the $D_1(2460)^0$ and $D_{2}^{*}(2460)^0$ mesons~\cite{PhysRevD.82.111101}. Using these values, our measurement of $\Gamma$, and the ratios from Ref.~\cite{PhysRevD.64.114004}, we calculate new values for the coupling constant $\hat{g}$.  Table~\ref{table:zach_eichten} shows the updated results.  We estimate the uncertainty on $\hat{g}$ assuming $\sigma_{\Gamma} \ll \Gamma$.  The updated widths reveal significant differences among the extracted values of $\hat{g}$. The order of magnitude increase in precision of the $D^{*+}$ width measurement compared to previous studies confirms the observed inconsistency between the measured $D^{*+}$ width and the chiral quark model calculation by Di Pierro and Eichten~\cite{PhysRevD.64.114004}. 

After completing this analysis, we became aware of Rosner's 1985 prediction that the $D^{*+}$ natural line width should be $83.9 \kev$ \cite{Rosner:1985dx}.  He calculated this assuming a  single quark transition model to use P-wave $K^* \to K\pi$ decays to predict P-wave $D^* \to D\pi$ decay properties. Although he did not report an error estimate for this calculation in that work, his central value falls well within our experimental precision. Using the same procedure and current measurements, the prediction becomes $(80.5 \pm 0.1) \kev$~\cite{rosnerPrivate}. A new lattice gauge calculation yielding  $\Gamma( D^{*+} ) = ( 76 \pm 7^{+8}_{-10} ) $ keV, has also been reported recently~\cite{Becirevic201394}.

\begin{table}
\begin{center}
\caption{Updated coupling constant values using the latest width measurements.  Ratios are taken from Ref.~\cite{PhysRevD.64.114004}.}
\begin{tabular}{c@{\hspace{3mm}}ccc}
\hline \hline \\[-1.7ex] 
\multirow{2}{*}{State} & \multirow{2}{*}{Width ($\Gamma$)} & $R = \Gamma/\hat{g}^2$ & \multirow{2}{*}{$\hat{g}$} \\
& & (model) & \\ \hline  \\[-1.7ex] 
$D^{*}(2010)^{+}$ & $83.3 \pm 1.3 \pm 1.4 \kev$ & $143 \kev$ & $0.76 \pm 0.01$ \\ 
$D_{1}(2420)^{0}$ & $31.4 \pm 0.5 \pm 1.3 \mev$ & $16 \mev$ & $1.40 \pm 0.03$ \\ 
$D_{2}^{*}(2460)^{0}$ &$50.5 \pm 0.6 \pm 0.7 \mev$ & $38 \mev$ & $1.15 \pm 0.01$ \\[-1.7ex]  \\ \hline \hline
\end{tabular}
\label{table:zach_eichten}
\end{center}
\end{table}

\input{acknow_PRL}

\bibliography{refs_bad2528}

\end{document}

%% file: authors_feb2013_bad2528.tex
%% author list as of 12-Feb-2013 (340 authors)
%
\author{J.~P.~Lees}
\author{V.~Poireau}
\author{V.~Tisserand}
\affiliation{Laboratoire d'Annecy-le-Vieux de Physique des Particules (LAPP), Universit\'e de Savoie, CNRS/IN2P3,  F-74941 Annecy-Le-Vieux, France}
\author{E.~Grauges}
\affiliation{Universitat de Barcelona, Facultat de Fisica, Departament ECM, E-08028 Barcelona, Spain }
\author{A.~Palano$^{ab}$ }
\affiliation{INFN Sezione di Bari$^{a}$; Dipartimento di Fisica, Universit\`a di Bari$^{b}$, I-70126 Bari, Italy }
\author{G.~Eigen}
\author{B.~Stugu}
\affiliation{University of Bergen, Institute of Physics, N-5007 Bergen, Norway }
\author{D.~N.~Brown}
\author{L.~T.~Kerth}
\author{Yu.~G.~Kolomensky}
\author{M.~J.~Lee}
\author{G.~Lynch}
\affiliation{Lawrence Berkeley National Laboratory and University of California, Berkeley, California 94720, USA }
% ...
\collaboration{The \babar\ Collaboration}
\noaffiliation

%% file: acknow_PRL.tex
We are grateful for the excellent luminosity and machine conditions
provided by our \pep2\ colleagues, 
and for the substantial dedicated effort from
the computing organizations that support \babar.
The collaborating institutions wish to thank 
SLAC for its support and kind hospitality. 
This work is supported by
DOE
and NSF (USA),
NSERC (Canada),
CEA and
CNRS-IN2P3
(France),
BMBF and DFG
(Germany),
INFN (Italy),
FOM (The Netherlands),
NFR (Norway),
MES (Russia),
MINECO (Spain),
STFC (United Kingdom). 
Individuals have received support from the
Marie Curie EIF (European Union)
and the A.~P.~Sloan Foundation (USA). 
The University of Cincinnati is gratefully acknowledged for its support of 
this research through a WISE (Women in Science and Engineering) fellowship to C. Fabby.